\documentclass[pre,showpacs,twocolumn,reprint,superscriptaddress]{revtex4-1}
\usepackage{graphicx,amsmath,amssymb,amsthm,braket}

\newcommand{\ee}{\mathrm{e}}
\newcommand{\ii}{\mathrm{i}}

\newcommand{\av}[1]{\overline{#1}}
\newcommand{\ham}[1]{\hat{H}^{(#1)}}
\newcommand{\avd}[1]{\langle#1\rangle_{\rm DE}}
\newcommand{\avme}[1]{\langle#1\rangle_{\rm MCE}}
\newcommand{\uf}{u_{f}}
\newcommand{\psii}{\psi_{\rm initial}}
\newcommand{\deff}{{d_{\rm eff}}}

\newcommand{\fl}{\delta O}
\newcommand{\ho}{\hat{O}}

\newcommand{\ci}[3]{#1\substack{+#2\\-#3}}

\begin{document}
\title{How accurately can the microcanonical ensemble describe\\ small isolated quantum systems?}
\author{Tatsuhiko N. Ikeda}
\affiliation{Department of Physics, University of Tokyo, Bunkyo-ku, Tokyo 113-0033, Japan}
\affiliation{Department of Physics, Harvard University, Cambridge, Massachusetts 02138, USA}
\author{Masahito Ueda}
\affiliation{Department of Physics, University of Tokyo, Bunkyo-ku, Tokyo 113-0033, Japan}
\affiliation{RIKEN Center for Emergent Matter Science (CEMS), Wako, Saitama 351-0198, Japan}

\date{\today}
\begin{abstract}
We numerically investigate quantum quenches of a nonintegrable hard-core Bose-Hubbard model to test the accuracy of the microcanonical ensemble in small isolated quantum systems. We show that, in a certain range of system size, the accuracy increases with the dimension of the Hilbert space $D$ as $1/D$. We ascribe this rapid improvement to the absence of correlations between many-body energy eigenstates. Outside of that range, the accuracy is found to scale as either $1/\sqrt{D}$ or algebraically with the system size.
\end{abstract}
\pacs{05.30.-d, 03.65.-w}
\maketitle
{\it Introduction.---}
The microcanonical ensemble (MCE)
is the fundamental working hypothesis in statistical mechanics
for the description of equilibrium states in an isolated system~\cite{landau1996statistical}.
However,
the range of applicability and the accuracy of the MCE have yet to be fully understood.
This problem dates back to von Neumann's seminal work~\cite{Neumann2010,Goldstein2010b,Goldstein2010a}
and has seen resurgence of interest recently~\cite{Polkovnikov2011a,Yukalov2011}
partly because
isolated quantum systems have been realized using ultracold atoms~\cite{Kinoshita2006,Trotzky2012,Gring2012,Langen2013}
and thus the verification of the hypothesis has now become an issue of practical importance as well.

The understanding of why the MCE can describe equilibrium states in the thermodynamic limit (TDL)
has deepened considerably in recent years.
Even under unitary time evolution,
an effective stationary state can appear
due to dephasing between many-body energy eigenstates~\cite{Tasaki1998,Reimann2008a,Linden2009,Short2011};
then, physical quantities are obtained by the weighted average
of expectation values over individual energy eigenstates (see Eq.~\eqref{eq:avd})
with the weights determined by the initial condition.
On the other hand, it is empirically known that the MCE,
which presupposes the equal-weighted average of those expectation values (see Eq.~\eqref{eq:avme}),
well describes the physical quantities in the effective stationary state,
regardless of the initial condition. 
Recent studies have suggested that
this success of the MCE derives from the fact
that quantum states with close eigenenergies typically emulate the same thermal state in the TDL.
This scenario has been formulated mathematically~\cite{Neumann2010,Goldstein2010b,Tasaki2010}
and verified by uniform random samplings of states
in a narrow energy shell~\cite{Sugita2006,Popescu2006,Goldstein2006,Reimann2007}.
The eigenstate thermalization hypothesis (ETH)~\cite{Deutsch1991,Srednicki1994,Rigol2008,Rigol2012}
is a quintessential representation of this scenario.
The ETH states that
expectation values of a macroscopic observable
for energy eigenstates
in a narrow energy shell, which we shall refer to as eigenstate expectation values (EEVs),
are equal in the TDL.
The ETH has been supported by several numerical studies
in various nonintegrable models~\cite{Rigol2009,Rigol2010,Santos2010a,Beugeling2014,Kim2014a,Steinigeweg2014,Khodja2015}.

While these studies concern
{\it why} the MCE works
in the TDL,
we address the question of
{\it how accurately} it works in small isolated quantum systems.
Such a question can now be tested experimentally~\cite{Serwane2011,Zurn2013,Wenz2013}.
We find that
the ETH gives, in small isolated quantum systems, an upper bound on the accuracy of the MCE
and the bound scales as $1/\sqrt{D}$,
where $D$ is the dimension of the Hilbert space~\cite{Beugeling2014,Steinigeweg2014}.
However, the accuracy itself has not yet been studied in concrete physical systems.

In this paper,
we numerically investigate the accuracy of the MCE for quantum quenches
in a one-dimensional nonintegrable tight-binding model of hard-core bosons (HCBs)
with the number of sites $L=15,18,21$ and $24$.
We show that,
for some quench magnitudes,
the accuracy improves proportionally to $1/D$,
which is much better than the upper bound given by the ETH mentioned above.
We argue that the $1/D$ scaling implies that
for quantum quenches in nonintegrable systems,
no correlations arise between many-body eigenstates.
We also show that as we increase the system size with the quench magnitude fixed,
three distinct regimes emerge in which the accuracy scales as
(i) $1/\sqrt{D}$ (ETH regime),
(ii) $1/D$ (no-correlation regime),
and (iii) $L^{-\alpha}$ with $\alpha>0$ (algebraic regime).

{\it Formulation of the problem.---}
We begin by formulating the problem in a general setup.
We consider an isolated quantum system described by
a time-independent Hamiltonian $\hat{H}$.
Let $\{\ket{E_n}\}_{n=1}^D$ be the set of the eigenstates of $\hat{H}$
with eigenenergies $\{E_n\}_{n=1}^D$:
$\hat{H}\ket{E_n}=E_n\ket{E_n}$ for $n=1,2,\dots,D$,
where $D(\gg1)$ is the dimension of the Hilbert space.
An initial pure state $\ket{\psii}$
evolves in time as
\begin{align}
\ket{\psi(t)}=\sum_{n=1}^D c_n \ee^{-\ii E_n t} \ket{E_n}\label{eq:expansion}
\end{align}
with $c_n \equiv \braket{n|\psii}$,
where the Planck constant is set to unity throughout this paper.
We assume that the energy gaps $\{E_n-E_m\}_{1\le m<n\le D}$ are all different~\cite{Tasaki1998,Reimann2008a}.
This assumption holds true in nonintegrable hard-core Bose-Hubbard models (see below)
or equivalent spin-1/2 systems (see {\it e.g.}, Ref.~\cite{Zangara2013}).
We also assume that an effective dimension~\cite{Short2011},
which is the effective number of the energy eigenstates involved in the initial state, is much greater than unity:
\begin{align}
d_{\rm eff}\equiv\frac{1}{\sum_n |c_n|^4}\gg1.\label{eq:effdim}
\end{align}
Then
the time-dependent expectation value of a few-body observable $\hat{O}$,
or $\braket{\psi(t)|\hat{O}|\psi(t)}=\sum_{m,n} c_n^* c_m \ee^{\ii(E_n-E_m)t}\braket{E_n|\hat{O}|E_m}$,
is most of the time close to its infinite time average,  since the off-diagonal ($m\neq n$) contributions cancel each other.
The infinite time average is represented by the diagonal ensemble average~\cite{Rigol2008}
\begin{align}
\avd{\hat{O}}\equiv \sum_{n=1}^D |c_n|^2 O_n,\label{eq:avd}
\end{align}
where $O_n \equiv \braket{E_n|\hat{O}|E_n}$ is called an eigenstate expectation value (EEV)
and $|c_n|^2$ represents the energy distribution.
Thus, Eq.~\eqref{eq:avd} describes the physical quantities in the effective stationary state.
We note that Eq.~\eqref{eq:avd} explicitly depends on the 
microscopic details of the initial state.

The MCE gives a different average $\avme{\ho}$ for the observable $\ho$,
which is defined, with
only two parameters, {\it i.e.,}
the central energy $E_C$ and the energy width $\delta$,
by
\begin{align}
\avme{\hat{O}} \equiv \mathcal{N}_{E_C,\delta}^{-1} \sum_{n\in I(E_C,\delta)} O_n,\label{eq:avme}
\end{align}
where $I(E_C,\delta)\equiv\{ n\, | \, E_n \in [E_C-\delta,E_C+\delta] \}$,
and the normalization factor $\mathcal{N}_{E_C,\delta} \equiv  \sum_{n\in I(E_C,\delta)}1$
gives the number of the energy eigenstates in the energy window $[E_C-\delta,E_C+\delta]$.
In this paper, we consider only those cases for which $\avme{\ho}$ coincides with $\avd{\ho}$ in the
TDL. Then we may define the error of the MCE due to the finite-size effect by
\begin{align}
\text{Error of MCE} \equiv  \avme{\ho} - \avd{\ho} \label{eq:error}
\end{align}
and numerically investigate this error in a concrete model.

{\it Model.---}
We use a one-dimensional model of HCBs
with the nearest- and next-nearest-neighbor hopping and interactions.
The Hamiltonian is given by
\begin{align}
\hat{H}^{(u)} = &\sum_{i=1}^L  \left[  -(\hat{b}_{i+1}^\dagger \hat{b}_i + \hat{b}_i^\dagger \hat{b}_{i+1})  + u \hat{n}_i \hat{n}_{i+1} \right]\notag\\
&+  \sum_{i=1}^L  \left[  - (\hat{b}_{i+2}^\dagger \hat{b}_i + \hat{b}_i^\dagger \hat{b}_{i+2})  + \hat{n}_i \hat{n}_{i+2} \right] ~, \label{eq:Hamiltonian}
\end{align}
where
the periodic boundary conditions are imposed and
 $\hat{b}_i$ ($\hat{b}_i^\dagger$) is the annihilation (creation) operator of a HCB on site $i$
with
$[\hat{b}_i,\hat{b}_j] = [\hat{b}_i^\dagger, \hat{b}_j^\dagger] = [\hat{b}_i,\hat{b}_j^\dagger ] = 0 \ \text{for}\ i\neq j$,
$\hat{b}^2=(\hat{b}^\dag)^2 = 0 \ \text{and} \  \{\hat{b}_i,\hat{b}_i^\dagger \} = 1$,
and $\hat{n}_i \equiv \hat{b}_i^\dagger \hat{b}_i$.
The total number $N$ of HCBs is conserved and chosen to be $N=L/3$ in our numerical study.

For simplicity,
the nearest- and next-nearest-neighbor hopping 
and the next-nearest-neighbor interaction energies are set to unity~\footnote{
We have confirmed that
the results shown in this paper
do not qualitatively change if
the next-nearest-neighbor hopping and interaction energies
are greater than $0.1$, where
the Wigner-Dyson-like statistics is seen.} in Eq.~\eqref{eq:Hamiltonian}.
In the following discussions, we consider quantum quenches
by suddenly changing the parameter $u$ from 0 to $\uf$.
In these quenches, $\ham{u}$ is nonintegrable
and the energy level spacings obey the Wigner-Dyson statistics~\cite{Bohigas1984}
due to the next-nearest-neighbor contributions~\cite{Santos2010}.

Due to the translational invariance of our model,
the Hilbert space is decomposed into $L$ sectors labeled by the translational momentum $P=2\pi m/L$ ($m=0,1,\dots,L-1$),
and we take the one with $P=2\pi/L$,
which cannot be decomposed into smaller sectors.
The dimension $D$ of this sector at each system size is shown in Table~\ref{tab:dims}
and the energy eigenstates in the sector are denoted by $\{\ket{E_n^{(u)}}\}_{n=1}^D$,
where the corresponding eigenenergies $\{E_n^{(u)}\}_{n=1}^D$ are arranged
in an ascending order.
All these eigenstates are obtained by numerically diagonalizing the Hamiltonian.

\begin{table}
\caption{
The number $N$ of HCBs,
the number $L$ of the lattice sites,
and the dimension $D$ of the Hilbert space for the sector with
the translational momentum $P=2\pi/L$
which is used in our numerical study.
}
\begin{tabular}{lrrrrr}\toprule 
$N$ &  5 & 6 & 7 & 8\\
$L$ &  15 & 18 & 21 & 24 \\
$D$ &  200 & 1026 & 5537& 30624 \\
\hline
\end{tabular}
\label{tab:dims}
\end{table}

{\it Protocol of our numerical experiment.---}
We consider a quantum quench where
the initial state $\ket{\psii}$ is
an eigenstate $\ket{E_{n_0}^{(0)}}$ of $\hat{H}^{(0)}$.
The time evolution is governed by $\hat{H}^{(\uf)}$
and an effective stationary state is eventually reached
where the expectation value of a few-body observable $\ho$
is given by the diagonal ensemble average~\eqref{eq:avd}
with $c_n=\braket{E_n^{(\uf)}|E_{n_0}^{(0)}}$.
Meanwhile, we calculate the MCE average of $\ho$~\eqref{eq:avme}
by setting $\delta=0.02L$ and $E_C$ so that
$\braket{\psii|\ham{\uf}|\psii}\approx \avme{\ham{\uf}}$.
Thus, we obtain the error of the MCE~\eqref{eq:error}
for the given initial state $\ket{E_{n_0}^{(0)}}$.

We calculate the errors of the MCE
starting from every eigenstate $\ket{E^{(0)}_{n_0}}$
whose ``effective inverse temperature'' $\beta_{n_0}$
falls in an interval $[0,0.05]$.
Here and henceforth, the Boltzmann constant is set to unity.
The effective inverse temperature $\beta_n$
of an eigenstate $\ket{E^{(0)}_n}$ is defined
by the equation $E_n^{(0)}=Z^{-1}\sum_{m=1}^D E_m^{(0)} \ee^{-\beta E_m^{(0)}}$,
where $Z\equiv\sum_{m=1}^D \ee^{-\beta E_m^{(0)}}$.
As shown in Table~\ref{tab:label_beta},
the eigenstates thus chosen lie in the middle of the spectrum~\footnote{
The initial states are also in the middle of the spectrum of $\ham{\uf}$.
The number of eigenstates whose eigenenergy is less than $\braket{\psi_{\rm initial}|\ham{\uf}|\psi_{\rm initial}}$
lies between $0.38D$ and $0.50D$ for any $\ket{\psi_{\rm initial}}$ considered in our study.
};
in this case
chaotic states appear~\cite{Brody1981,Zelevinsky1996}
and
thermalization occurs~\cite{Santos2010,Rigol2010,Santos2010a,Torres-Herrera2013}.

\begin{table}
\caption{
The label $n$ of the eigenstate
that has the effective temperature closest to $\beta=0.05$ and $0.00$.
The ratio $n/D$ shows where each eigenstate lies in the spectrum.
}
\newlength{\myheight}
\setlength{\myheight}{1cm}
\begin{tabular}{lcrrrrr}\toprule 
\multicolumn{2}{c}{$L$} & 15 & 18 & 21 & 24 \\ \hline
{$\beta=0.05$}& $n$  & 79 & 403 & 2140 & 11696\\ 
& $n/D$  & 0.40 & 0.39 & 0.39 & 0.38\\ \hline
$\beta=0.00$& $n$  & 96 & 485 & 2628 & 14578\\
 & $n/D$  & 0.48 & 0.47 & 0.47 & 0.48\\
\hline
\end{tabular}
\label{tab:label_beta}
\end{table}

We investigate two local operators $\hat{O}_1\equiv\hat{n}_1\hat{n}_2$
and $\hat{O}_2\equiv \hat{n}_1\hat{n}_3$
representing
the correlations
in the numbers of HCBs
between the nearest and next-nearest neighbors, respectively.
We have confirmed that
our results shown in the following are qualitatively unaltered
for other two local operators,
$b_1^\dag b_2+{\rm H.c.}$ and $b_1^\dag b_3+{\rm H.c.}$.
We note $|\avd{\hat{O}}-\avme{\hat{O}}|\le1$ for $\hat{O}=\hat{O}_1$ and $\hat{O}_2$
because their operator norms are unity.

\begin{figure}
\begin{center}
\includegraphics[width=8.5cm]{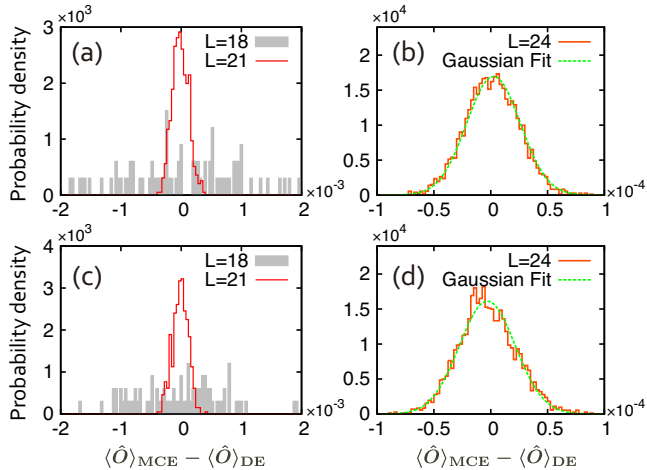}
\caption{(Color Online)
The distribution of the errors of the MCE (Eq.~\eqref{eq:error}) obtained for each of the initial states
in the quench of $\uf=0.4$ 
for (a) [(c)] $L=18$ (filled box) and 21 (solid line) and (b) [(d)] $L=24$ (solid line) for $\hat{O}=\hat{O}_1$ [$\ho_2$].
The dashed curves in (b) and (d) are the least squares fits
of the distributions with the Gaussian distributions.
We note that the horizontal and vertical axes are scaled by
the factors shown at the right-bottom and left-top corners of the panels, respectively.
}
\label{fig:Od-Ome}
\end{center}
\end{figure}

{\it The $1/D$ scaling of the accuracy.---}
Figure~\ref{fig:Od-Ome} illustrates
the distribution of $\avd{\hat{O}}-\avme{\hat{O}}$ obtained for each initial state
in the quench of $\uf=0.4$.
Figures~\ref{fig:Od-Ome}(a) and (b)
show that, for $\ho_1$,
the errors become markedly concentrated near zero as the system size increases.
Figures~\ref{fig:Od-Ome}(c) and (d)
show that the distribution of the errors for $\hat{O}_2$ behaves similarly to that for $\hat{O}_1$.
We discuss, in the following, how fast the width of the distribution vanishes
by examining the root mean square (RMS) of the errors,
which we call the accuracy of the MCE.

\begin{figure}
\begin{center}
\includegraphics[width=8.5cm]{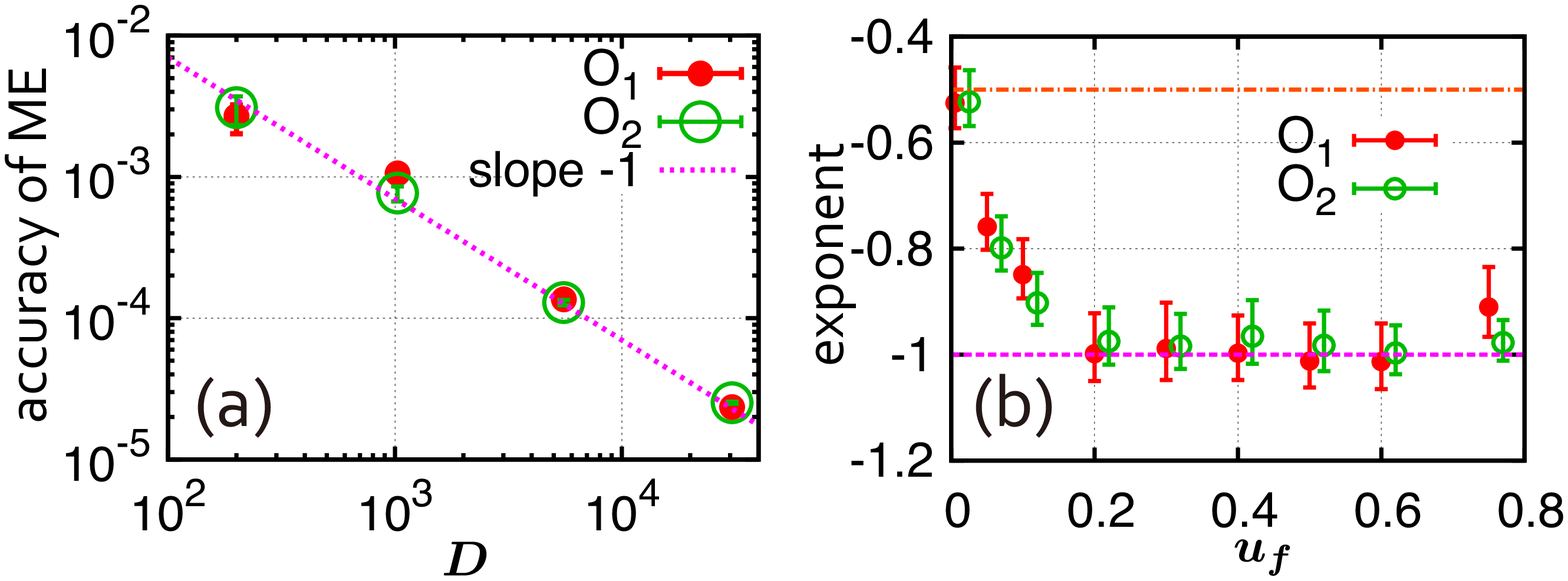}
\caption{(Color Online)
(a) Accuracy of the MCE
for two local operators $\hat{O}=\hat{O}_1$ (circles) and $\hat{O}_2$ (triangles)
at $L=15,18,21$ and $24$ (see Table~\ref{tab:dims} for the corresponding $D$)
with $\uf=0.4$.
The error bars show the estimation errors~\cite{commentesterr}.
The data points are well fitted by the dotted line which has the slope of $-1$.
(b)
For $\hat{O}_1$ (circles) and $\hat{O}_2$ (triangles, slightly shifted to the right for clarity),
the exponent $B$
obtained by the least squares fits of the accuracy with a function $f(D)=AD^B$,
where the error bars represent the 95\% confidence intervals.
The dash-dotted line indicates the value of $-1/2$ as predicted by the ETH.
}
\label{fig:D-1}
\end{center}
\end{figure}

The first main result of this paper is that,
in the quench of $\uf=0.4$,
the accuracy
is proportional to $1/D$ for both $\hat{O}_1$ and $\hat{O}_2$
as shown in Fig.~\ref{fig:D-1}(a),
where the accuracy is plotted against $D$
with the error bars representing the estimation errors~\cite{commentesterr}.
By conducting the least squares fits
of the accuracy with a function $f(D)=AD^B$,
we find the exponent $B$ to be
$\ci{-1.00}{0.07}{0.05}$ for $\ho=\ho_1$
and $\ci{-0.97}{0.08}{0.06}$ for $\ho=\ho_2$
with 95\% confidence
($\ci{-0.99}{0.06}{0.05}$ 
for $\ho=b_1^\dag b_2+{\rm H.c.}$ and $\ci{-0.97}{0.10}{0.07}$ for $\ho=b_1^\dag b_3+{\rm H.c.}$),
which clearly demonstrates the $1/D$ scaling of the accuracy.
We have also conducted similar analyses for
various quench magnitudes
and found the $1/D$ scaling for the range of $0.2\leq \uf < 0.75$
as shown in Fig.~\ref{fig:D-1}(b).
The discrepancies of the exponents from $-1$
seen for $0\le\uf\le0.2$ and $\uf\ge0.75$ 
will be addressed later,
and we here focus on the implications of the $1/D$ scaling of the accuracy.

We discuss the $1/D$ scaling in terms of
the number $N$ of HCBs in our model
with a general filling factor $\nu\equiv{}N/L$,
which has been fixed to be $1/3$ in the above discussions.
The dimension $D$ is approximately given by
$D\approx\binom{L}{N}/L\approx10^{\gamma(\nu)N}$,
where $\gamma(\nu)=-\nu^{-1}[\nu \log_{10}\nu + (1-\nu)\log_{10}(1-\nu)]$
is a monotonically decreasing function
giving, in particular, $\gamma(1/4)=0.9768\dots \approx 1$.
Thus the $1/D$ scaling implies that, at 1/4 filling,
the accuracy improves by one order of magnitude
as we increase the number of HCBs by one.

{\it Two ETH upper bounds on the accuracy.---}
Before discussing the underlying mechanism for the $1/D$ scaling of the accuracy,
we derive from the ETH two upper bounds on the accuracy of the MCE
by assuming that the energy distribution $|c_n|^2$ is localized,
or its width is much smaller than the macroscopic energy scale such as the total energy.
This assumption is needed for the state to be thermodynamically normal
in the sense that the total energy is macroscopically definite.
Then we point out that the upper bounds decrease proportionally to $1/\sqrt{D}$,
implying that the ETH alone cannot explain the $1/D$ scaling.

First, we note that the EEVs $O_n$ of a few-body observable $\hat{O}$
in nonintegrable systems
are known to behave as
\begin{align}
O_n = f(E_n/L) + \fl_n,\label{eq:Oo}
\end{align}
where $f(x)$ is a smooth function
and $\fl_n$ represents random fluctuations around it~\cite{Srednicki1996}.
Correspondingly,
the error of the MCE
is decomposed into two parts:
\begin{align}
\avd{\hat{O}}-\avme{\hat{O}}=\Delta O_\text{sys} +\Delta O_\text{rand},\label{eq:sysrand}
\end{align}
where
\begin{align}
\Delta O_\text{sys}&\equiv \sum_n |c_n|^2 f(E_n/L)
-f_{\rm MCE},\label{eq:Osys}\\
\Delta O_\text{rand}&\equiv \sum_n |c_n|^2 \delta O_n
-\delta O_{\rm MCE}. \label{eq:Orand}
\end{align}
with $f_{\rm MCE}\equiv \mathcal{N}_{E_C,\delta}^{-1} \sum_{n\in I(E_C,\delta)}f(E_n/L)$
and $\delta O_{\rm MCE}\equiv \mathcal{N}_{E_C,\delta}^{-1} \sum_{n\in I(E_C,\delta)}\fl_n$.

Second, we note that $\Delta O_\text{sys}$ is negligible
if the energy distribution $|c_n|^2$ is sufficiently localized.
In fact, under this condition,
the Taylor expansion of $f(E_n/L)$ up to the first order is sufficient
and we have $\Delta O_\text{sys}\approx0$
because we choose $E_C$ so that $\braket{\psii|\ham{\uf}|\psii}\approx\avme{\ham{\uf}}$.
We discuss the higher order contributions in Discussion below.

Thus,
if the energy distribution $|c_n|^2$ is sufficiently localized,
the error of the MCE is dominated by $\Delta O_\text{rand}$,
which is bounded from above strictly by $\Delta_O\equiv 2\max_n |\fl_n|$
and roughly by the standard deviation $\sigma_O$ of $\{\fl_n\}_n$.
We note that $\Delta_O$ and $\sigma_O$ should be calculated
in an energy window where $|c_n|^2$'s are significantly weighted
instead of the entire spectrum.
In the following discussion, we take the window to be the microcanonical one
because the width $|c_n|^2$ has turned out to be smaller than $\delta$ for $\uf < 0.75$,
where we discuss the $1/D$ scaling.
We call both $\Delta_O$ and $\sigma_O$ the ETH upper bounds
because they are commonly used as the indicators of the ETH
in the strong and weak senses, respectively,
which imply $\Delta_O\to0$ and $\sigma_O\to0$ in the TDL~\cite{Biroli2010,Ikeda2013,Kim2014a,Beugeling2014,Steinigeweg2014,Khodja2015}.

\begin{figure}
\begin{center}
\includegraphics[width=8.5cm]{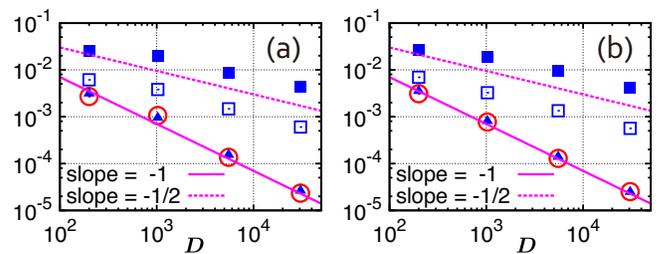}
\caption{(Color Online)
Accuracy of the MCE (circles),
strong (filled squares, $\Delta_O$)
and weak (open squares, $\sigma_O$) ETH indicators,
the indicator for the no-correlation model (triangles, $\tilde{\sigma}_O$ defined in Eq.~\eqref{eq:no_cor})
for (a) $\hat{O}_1$ and (b) $\hat{O}_2$ with $\uf=0.4$.
The solid and dashed lines with slopes $-1$ and $-1/2$, respectively,
are guides to the eye.
The numerically obtained accuracy is best fitted by the no-correlation model.
}
\label{fig:scenarios}
\end{center}
\end{figure}

It has recently been shown that $\sigma_O$ approaches zero
proportionally to $1/\sqrt{D}$ in nonintegrable spin systems~\cite{Beugeling2014,Steinigeweg2014,Khodja2015}.
We have also obtained the $1/\sqrt{D}$ scaling for both ETH indicators $\Delta_O$ and $\sigma_O$
as illustrated in Fig.~\ref{fig:scenarios}.
These results imply that
the ETH alone cannot explain the $1/D$ scaling found in our numerical study.

{\it The underlying mechanism for the $1/D$ scaling.---}
The second main result is that
yet another indicator
\begin{align}
\tilde{\sigma}_o \equiv \frac{\sigma_o}{\sqrt{d_{\rm eff}}}\label{eq:no_cor}
\end{align}
can describe the accuracy as illustrated in Fig.~\ref{fig:scenarios}.
This indicator involves, in addition to $\sigma_O$,
an extra suppression factor $1/\sqrt{d_{\rm eff}}$.
Since $d_{\rm eff}$ represents the effective number of nonzero terms
on the RHS of Eq.~\eqref{eq:Orand},
the extra suppression factor $1/\sqrt{d_{\rm eff}}$
implies that there is little correlation between the terms.

It has been proposed as an alternative to the ETH~\cite{Peres1984,Ikeda2011,Rigol2008,Rigol2012} that
the absence of correlation between
the energy distribution $|c_n|^2$
and the EEV $O_n$
suppresses the error of the MCE.
Indeed this is the case for our setup
since the initial states are chosen independently of 
the observables $\hat{O}_1$ and $\hat{O}_2$.
However, this mechanism alone cannot explain why
the accuracy is so good as
Eq.~\eqref{eq:no_cor}
if $|c_n|^2$'s are correlated with each other.

Let us derive Eq.~\eqref{eq:no_cor} by
introducing the ``no-correlation model''
in which there are no correlations between $|c_n|^2$'s except for the constraints $\sum_n |c_n|^2=1$
and $\sum_n |c_n|^4=1/\deff$.
To be more specific, it is given by $|c_n|^2=q_n x_n$,
where $q_n$ represents a smooth profile of $|c_n|^2$ satisfying $\sum_n q_n=1$, $\sum_n q_n^2 = \deff^{-1}$, and $q_n=0$ for $n\not\in I(E_C,\delta)$,
and $\{x_n\}_{n=1}^D$ are positive random variables whose mean and standard deviation are both unity.
These conditions are known to hold
in the middle of the spectrum in chaotic systems
with $q_n$ being Gaussian~\cite{Santos2012p,Santos2012q,Torres-Herrera2013}.
The no-correlation model leads to
$\overline{(\Delta O_\text{rand})^2}=
 (\sigma_O^{r})^2/\deff + (\delta O^q - \delta O_{\rm MCE})^2$,
where
$\delta O^{q}\equiv\sum_n q_n \delta O_n$, and $(\sigma_O^{r})^2=\sum_n r_n (\delta O_n)^2$ with $r_n=q_n^2/\sum_m q_m^2$.
Here  $\av{\cdots}$ denotes the statistical average with the probability distribution $P(x_1,x_2,\dots,x_D)\equiv p(x_1)p(x_2)\dots p(x_D)$.
Since $\{q_n\}_n$ and $\{r_n\}_n$ are smooth and normalized profiles within the microcanonical window,
we assume $\delta O^q=\delta O_{\rm MCE}$ and $(\sigma_O^r)^2=\sigma_O^2$.
Then, we obtain $\overline{(\Delta O_\text{rand})^2}=\sigma_O^2/\deff$,
which implies that Eq.~\eqref{eq:no_cor} gives an estimation for $\Delta O_{\rm rand}$.

Our result that the no-correlation model explains the $1/D$ scaling of the accuracy
implies that
we cannot induce correlations between $|c_n|^2$'s through a sudden change of a single parameter in the middle of the spectrum where many-body energy eigenstates show chaotic behavior.
While we can control the total energy by changing the parameter,
we cannot manipulate individual many-body eigenstates
whose landscape changes drastically from one to the neighboring one for nonintegrable systems.

{\it Discussions.---}
First, we discuss how the $1/D$ scaling is modified in larger system sizes.
As we increase $L$ further,
there might exist an $L_\text{upper}$ at which $\Delta O_\text{sys}$ becomes comparable with $\Delta O_\text{rand}$
and can no longer be ignored.
This is because, whereas $\Delta O_{\rm rand}$ decays as $1/D$,
$\Delta O_{\rm sys}$ decays only algebraically with $L$ for thermodynamically normal states as follows.
By making the Taylor expansion of $f(E_n/L)$,
we have $\Delta O_{\rm sys}\sim 2^{-1}f''(E_C/L)[\Delta E_{\rm DE}^2-\Delta E_{\rm MCE}^2]/L^2 $
because the zeroth- and first-order contributions cancel out due to the normalization condition
and the fact that $E_C$ is chosen to be equal to $\braket{\psii|\ham{\uf}|\psii}$.
Here $\Delta E_{\rm DE}^2$ and $\Delta E_{\rm MCE}^2$ denote those the energy fluctuations in the DE and MCE, respectively, that are proportional to $L$, and we obtain $\Delta O_{\rm sys} \propto L^{-1}$.
Even if we perform the fine-tuning of $\delta$ so that $\Delta E_{\rm DE}^2=\Delta E_{\rm MCE}^2$,
the contributions from the Taylor expansion of $f(E_n/L)$ at all orders cannot be canceled in general,
and $\Delta O_{\rm sys}$ decreases only algebraically with $L$.

\begin{figure}
\begin{center}
\includegraphics[width=8.5cm]{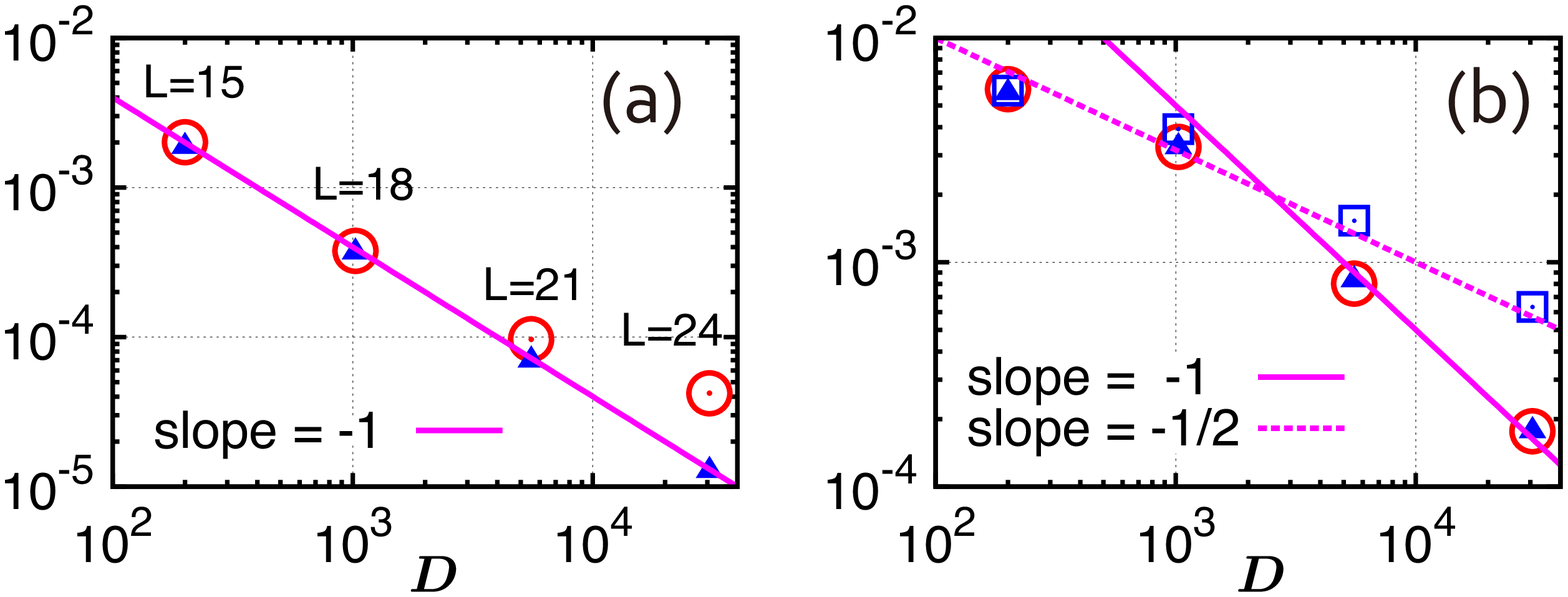}
\caption{(Color Online)
Accuracy of the MCE (circles)
and indicators for the no-correlation model (triangles, $\tilde{\sigma}_O$ defined in Eq.~\eqref{eq:no_cor})
and the weak ETH (open squares, $\sigma_O$)
for $\hat{O}_1$ with (a) $\uf=1$ and (b) $\uf=0.05$.
The solid and dashed lines with slopes $-1$ and $-1/2$, respectively,
are guides to the eye.
}
\label{fig:exceptions}
\end{center}
\end{figure}

The $L_\text{upper}$ should be smaller when
the energy distribution $|c_n|^2$ is less localized
and the higher-order contributions in the Taylor expansion of $f(E_n/L)$ become important.
In fact, for the large quench with $\uf=1$,
the deviation from the $1/D$ scaling is seen at $L=21$ and $24$ as shown in Fig.~\ref{fig:exceptions}(a).
The deviation may be interpreted to be the cause
for the deviation from the $1/D$ scaling at the largest quench in Fig.~\ref{fig:D-1}(b).

Second, we discuss how the $1/D$ scaling changes in smaller system sizes.
As we decrease the system size with the quench magnitude held fixed,
the quench energy becomes smaller than the energy level spacings
and we enter the regime where $\deff\sim1$.
Thus, for a given quench magnitude, there exists $L_\text{lower}$
below which the $1/D$ scaling disappears.
In this case, Eq.~\eqref{eq:no_cor} reduces to the indicator of the ETH in the weak sense
and the accuracy is proportional to $1/\sqrt{D}$.
This crossover between the $1/D$ scaling and the $1/\sqrt{D}$ scaling
is seen for $\uf=0.05$ as shown in Fig.~\ref{fig:exceptions}(b)~\footnote{
The data (not shown) for $L=12$ are consistent with this observation.
}.
Thus, the lack of mixing of numerous energy eigenstates
is the cause for the crossover from the $1/D$ to $1/\sqrt{D}$ scalings for the small quenches as can be seen for $\uf<0.2$ in Fig.~\ref{fig:D-1}(b).

Thus, 
we find three regimes of the system size for a given quench magnitude:
(i) $L<L_\text{lower}$ where the accuracy is described by the ETH and proportional to $1/\sqrt{D}$,
(ii) $L_\text{lower}<L<L_\text{upper}$ where the accuracy is proportional to $1/D$ due to the absence of
correlations between the many-body eigenstates,
and (iii) $L_\text{upper}<L$ where the accuracy improves only algebraically with $L$
because $\Delta O_\text{sys}$
rather than $\Delta O_\text{rand}$
dominates in the error of the MCE (see Eq.~\eqref{eq:sysrand}).

Our findings imply that
there exists an exponential enhancement of the accuracy
in the small system sizes
and the MCE can describe equilibrium states very accurately even in small isolated quantum systems.
In fact, 
the accuracy of the MCE reaches $10^{-4}$ or even better
in the system with only 8 HCBs on 24 sites
as shown in Figs.~\ref{fig:scenarios} and \ref{fig:exceptions}.

{\it Conclusions.---}
We have numerically investigated the accuracy of the MCE
in interaction quenches for a nonintegrable hard-core Bose-Hubbard model.
We have found a regime where the accuracy improves proportionally to $1/D$ (see Fig.~\ref{fig:D-1}).
This rapid improvement of the accuracy
implies that quenching a single parameter cannot induce
correlations between the numerous many-body eigenstates
since they depend nontrivially on the parameter in nonintegrable systems.
As we increase the system size,
there are three regimes where the accuracy scales as
(i) $1/\sqrt{D}$ (ETH regime),
(ii) $1/D$ (no-correlation regime),
and
(iii) algebraically with $L$.
Due to the regimes (i) and (ii), where
the accuracy improves exponentially with $L$,
the MCE can describe the equilibrium states 
quite accurately even in small systems.

{\it Acknowledgements.---} 
Fruitful discussions with Shunsuke Furukawa, Kohaku H.~Z.~So, and Tomohiro Shitara are gratefully acknowledged.
We also thank Hyungwon Kim to have given helpful comments
on the manuscript.
This work was supported by
KAKENHI 26287088, 
a Grant-in-Aid for Scientific Research on Innovation Areas ``Topological Quantum Phenomena'' (KAKENHI 22103005),
and the Photon Frontier Network Program,
from MEXT of Japan.
T.N.I. acknowledges the JSPS for
financial support (Grant No. 248408)
and Postdoctoral Fellowship for Research Abroad..

\bibliography{manuscript_PRE_resubmit}

\begin{thebibliography}{50}
\bibitem [{com()}]{commentesterr}%
  \BibitemOpen
  \href@noop {} {}\bibinfo {note} {We regard the mean square of our data as the
  sum of the squared mean and the variance, which are unbiasedly estimated.
  Then, with the chain rule, we obtain the error of the mean square from those
  of the mean and the variance, which are estimated from the variances of their
  unbiased estimators.}\BibitemShut {Stop}%
\end{thebibliography}
\end{document}